\documentclass[aps,pra,showpacs,floatfix,superscriptaddress]{revtex4}
\usepackage{graphicx}
\usepackage{color}
\usepackage{amssymb}
\usepackage{subfigure}
\usepackage{color}
\usepackage{ulem}

\newtheorem{thm}{Theorem}

\begin{document}
\title{Complementarity in atomic and oscillator systems}
\author{R. Srikanth}
\email{srik@ppisr.res.in}
\affiliation{Poornaprajna Institute of Scientific Research, Sadashiva
Nagar, Bangalore- 560 080, India.}
\affiliation{Raman Research Institute, Sadashiva Nagar,
Bangalore - 560 080, India\\ Phone: +91 9844 593440}
\author{Subhashish Banerjee}
\email{subhashish@cmi.ac.in}
\affiliation{Chennai Mathematical Institute, Padur PO, Siruseri 603103, India}

\begin{abstract}
  We develop  a unified,  information theoretic interpretation  of the
  number-phase   complementarity   that    is   applicable   both   to
  finite-dimensional  (atomic)  and infinite-dimensional  (oscillator)
  systems.
The relevant uncertainty principle is obtained  as a lower bound on {\it
    entropy excess}, the difference between number entropy and phase
knowledge, the latter defined as the relative entropy with respect to
the uniform distribution. 
\end{abstract}

\maketitle


Two  observables  $A$  and  $B$  of  a  $d$-level  system  are  called
complementary  if  knowledge of  the  measured  value  of $A$  implies
maximal  uncertainty of  the measured  value  of $B$,  and vice  versa
\cite{mu88}.   Complementarity is  an aspect  of  the Heisenberg
uncertainty  principle, which  says  that for  any  state $\psi$,  the
probability  distributions obtained  by measuring  $A$ and  $B$ cannot
both  be   arbitrarily  peaked  if   $A$  and  $B$   are  sufficiently
non-commuting.  Expressed in terms of measurement entropy 
the Heisenberg  uncertainty principle takes the form:
\begin{equation}
H(A) + H(B) \ge \log d.
\label{eq:hu0}
\end{equation}
where $H(A)$  and $H(B)$  are the Shannon  entropy of  the measurement
outcomes   of  a  $d$-level   quantum  system   \cite{nc00,delg}.  Eq.
(\ref{eq:hu0}) has several advantages over the traditional uncertainty
multiplicative form \cite{kraus,mu88,deu83,par83}.

More   generally,    given   two   observables    $A   \equiv   \sum_a
a|a\rangle\langle a|$ and $B  \equiv \sum_b b|b\rangle\langle b|$, let
the  entropy   generated  by   measuring  $A$  or   $B$  on   a  state
$|\psi\rangle$  be given  by,  respectively, $H(A)$  and $H(B)$.   The
information  theoretic representation  of  the Heisenberg  uncertainty
principle      states       that      $H(A)      +       H(B)      \ge
2\log\left(\frac{1}{f(A,B)}\right)$,       where       $f(A,B)       =
\max_{a,b}|\langle a|b\rangle|$, and  $H(\cdot)$ is the Shannon binary
entropy.    A  pair   of   observables,  $A$   and   $B$,  for   which
$f(A,B)=d^{-1/2}$  are  said to  form  mutually  unbiased bases  (MUB)
\cite{ii81,dur05}.   Conventionally,  two  Hermitian  observables  are
called complementary only if they are mutually unbiased.

An application of this idea to obtain an entropic uncertainty relation
for oscillator systems in the Pegg-Barnett scheme \cite{pb89} has been
made in Ref. \cite{abe}, and for higher entropic uncertainty relations
in  Ref.  \cite{wiwe}.  An  algebraic  treatment  of  the  uncertainty
relations,  in  terms  of  complementary subalgebras,  is  studied  in
Ref. \cite{petz}.

An extension of Eq. (\ref{eq:hu0}) to the case where $A$ or $B$ is not
discrete is considered in Ref. \cite{rs07}, where the problem that the
Shannon entropy  of a  continuous random variable  may be  negative is
circumvented   by  instead   using  relative   entropy   (also  called
Kullb{\"a}ck-Leibler divergence, which is always positive) \cite{kl51,
  hj03} with  respect to a  uniform distribution.  This quantity  is a
measure  of knowledge  \cite{rs07}.   Note that  recourse to  entropic
knowledge may not  be always necessary, and other  ways might exist to
circumvent the problem.  In the case of position $X$ and momentum $P$,
the sum of  functionals, $H[X] + H[P]$ is  always positive even though
the  summands  may  be  negative  \cite{beck75}.  Further,  one  might
consider exponentiating the entropy  to ensure positivity. But neither
of these options is, in our opinion, intuitive from the perspective of
interpretation.  Finally, it  may well turn out that  for all physical
states of  a system,  the continuous variables  in question  may never
yield negative  entropies. In  Ref. \cite{rs07}, we  numerically found
this  to be the  case for  the continued-valued  phase obervable  in a
two-level system. However,  we know of no (published)  proof that this
is  true in  general, which  was  the motivation  behind adopting  the
concept of entropic knowledge in the uncertainty relation.

An   example  where  our   re-expressed  entropic   uncertainty  finds
application would be when one of the observables, say $A$, is bounded,
and its conjugate $B$ is described  not as a Hermitian operator but as
a {\it  continuous-valued} POVM. (There are  no continuous observables
corresponding to  {\it projective measurements}  in finite-dimensional
Hilbert spaces.)  A  particular case of discrete-continuous conjugacy,
considered in detail  in Ref. \cite{rs07}, is the  number and phase of
an  atomic system.   This generalization  of the  entropic uncertainty
principle to cover discrete-continuous  systems still suffers from the
restriction that the  system must be finite dimensional,  since in the
case  of  an  infinite-dimensional  system,  such  as  an  oscillator,
entropic knowledge  of the number distribution can  diverge, making it
unsuitable for  infinite-dimensional systems.  Therefore to  set up an
entropic  version of the  uncertainty principle,  that unifies  and is
applicable  to  all  systems,  including infinite  dimensional  and/or
continuous-variable  systems,   it  may  be  advantageous   to  use  a
combination of  entropy and  knowledge, in particular,  the difference
between entropy of the discrete, infinite observable and between phase
knowledge.  This is discussed in detail below.


The quantum  description of phases  \cite{pp98} has a long  history of
\cite{pb89,pad27,   sg64,   cn68,ssw90,   scr93};   see   also   Refs.
\cite{ssw91, mh91}.  In the quantum  theory of detection, the issue of
quant  phases appears  quite fundamentally  as a  lack  of phase-shift
self-adjoint  operator that is covariant under shifts generated by
the number observable  \cite{holevo,pel91}.  In  a  recent
approach, which  we adopt, the  concept of phase distribution  for the
quantum phase has been introduced \cite{ssw91,sb06,as92,as96}. 
This distribution is the expectation value of the
{\it canonical phase}, the POVM obtained by setting all entries
in the phase matrix (defining the generalized phase POVM) 
equal to 1 \cite{bu01}. This reference also  discusses different 
forms of complementarity of number and phase, depending on the defintion
of phase chosen. In particular, number and canonical phase are not
complementary, but satisfy the weaker condition of 
{\it value complementarity}, as defined there. 
Our present work can be considered as an entropic interpretation of 
the value complementarity of number and canonical phase.
In this section, we
briefly recapitulate, for convenience, some useful formulas of quantum
phase  distributions for oscillator  systems. For  the case  of atomic
systems, the basic formulas were presented in \cite{rs07}.

We define a phase distribution  ${\cal P}(\theta)$ for a given density
operator  $\rho$, which  in  our  case would  be  the reduced  density
matrix, as
\begin{equation}
{\cal P}(\theta) = {1 \over 2\pi} \langle \theta|\rho| \theta 
\rangle = {1 \over 
2\pi} \sum\limits_{m, n=0}^{\infty} \rho_{m, n} e^{ i(n-
m)\theta}, ~ 0 \leq \theta \leq 2\pi  \label{osph} 
\end{equation}
where  the   states  $|\theta\rangle$  are  the   eigenstates  of  the
Susskind-Glogower   \cite{sg64}   phase   operator  corresponding   to
eigenvalues of unit  magnitude and are defined in  terms of the number
states  $|n\rangle$  as  $|\theta\rangle =  \sum\limits_{n=0}^{\infty}
e^{i n \theta} |n\rangle$, and $|\theta\rangle\langle\theta|$ is the
canonical phase POVM.  The sum in Eq.  (\ref{osph}) is assumed to
converge.  The phase distribution  is positive definite and normalized
to unity with $\int_\theta |\theta\rangle\langle\theta|d\theta=1$.

The complementary number distribution is
\begin{equation}
p(m) = \langle m|\rho| m \rangle, \label{osnu}
\end{equation}
where  $|m \rangle$  is the  number (Fock)  state.   Analogous results
exist for atomic states, with the Susskind-Glagower states replaced by
atomic  coherent  states  \cite{mr78,  ap90},  and  number  states  by
Wigner-Dicke states \cite{at72,rd54}.


Defining  entropic knowledge  $R[f]$  of random  variable  $f$ as  its
relative   entropy   with   respect   to  the   uniform   distribution
$\frac{1}{d}$,  i.e., $R[f]  \equiv  S\left(f(j)||\frac{1}{d}\right) =
\sum_j   f(j)\log(df(j))$,  we   can  recast   Heisenberg  uncertainty
principle in terms  of entropy $H$ and knowledge $R$,  as shown by the
theorem:
\begin{thm}
  Given two Hermitian observables $A$ and $B$ that form a pair of MUB
in a finite dimensional Hilbert space,
  the uncertainty relation (\ref{eq:hu0}) can be expressed as
\begin{equation}
X(A,B) \equiv H(A) - R(B) \ge 0.
\label{eq:ra}
\end{equation}
\end{thm}
{\bf Proof.} Let the distribution obtained by measuring $A$ and $B$ on
a given state be, respectively, $\{p_j\}$ and $\{q_k\}$.  Denoting
$H(A) \equiv -\sum_j p_j 
\log_2 p_j$, the l.h.s of Eq. (\ref{eq:ra}) is
given by
$H\left(A\right) - S\left(B||\frac{1}{d}\right) =
H(A) - \sum_k q_k \log( dq_k) 
= H(A) + H(B) - \log d \label{eq:sub} \\
\ge 2\log\left(\frac{1}{f(A,B)}\right) - \log d$,
where the last equation follows from Ref. \cite{mu88}.
For   a   pair   of  MUB  \cite{kraus,deu83},
$f(A,B)=d^{-1/2}$, from which the theorem follows.  \hfill $\blacksquare$
\bigskip \\ 

Phyically Eq. (\ref{eq:ra}) expresses that ignorance of one of two MUB
variables is  at least as large  as the knowledge of  the other.  Some
properties of  $X$ are evident.  Clearly, $X(A,B)=X(B,A)$.   It is not
difficult to see that $X(A,B)$  attains its largest value of $\log(d)$
when $A$ and $B$  are MUBs (and the state is an  eigenstate of a third
MUB),  and  its minimum  value  of $-\log(d)$  when  $A$  and $B$  are
commuting (and the  state is an eigenstate of  either observable).  We
may  quantify the `degree  of complementarity'  in the  following way.
Since  the  above bound  is  tight,  we  define $X_{\min}(A,B)  \equiv
2\log\left(\frac{1}{f(A,B)}\right) - \log d$  as the smallest value of
$X(A,B)$  over all  possible  states  for a  given  pair of  Hermitian
observables $A$  and $B$. This is a  monotonically decreasing function
in  the interval  $[d^{-1/2},1]$,  going from  0  to $-\log(d)$.   Two
Hermitian observables  $A$ and $B$ are  maximally complementary (i.e.,
form  an  MUB) if  $X_{\min}(A,B)=  0$.   are minimally  complementary
(i.e., are compatibe) if $X_{\min}(A,B)=-\log(d)$.

A point  worth noting about Eq.  (\ref{eq:ra}) is that  it contains no
explicit mention of dimension $d$.  What is remarkable is that we find
this situation persists  even when one of $A$ or  $B$ is not discrete,
but  a continuous-valued  POVM  (for discrete-valued  POVMs, cf.  Ref.
\cite{mas07}),  and  furthermore,  the  system  is  no  longer  finite
dimensional  but  instead infinite  dimensional.  The only  additional
requirement is  that the continuous-valued  variable should be  set as
$B$ (the knowledge- rather  than the ignorance-variable), since $H(B)$
can potentially  be negative for  such variables.  This  makes $X(A,B)
\ge 0$  as a  very succinct and  general statement of  the uncertainty
principle.   By contrast,  because there  is no  prior  guarantee that
measurement   entropy   $H(\cdot)$   will   be  non-negative   for   a
continuous-valued observable,  it is not  obvious that the  version of
the  Heisenberg  uncertainty  principle  given  by  (\ref{eq:hu0})  is
generally  applicable,  and furthermore,  because  there  is no  prior
guarantee  that  measurement  entropic  knowledge $R(\cdot)$  will  be
well-defined for  infinite-dimensional variables, the  version $R(A) +
R(B) \le \log(d)$ of Ref.  \cite{rs07} is also not obviously generally
applicable.

One catch is that on  account of the POVM-nature \cite{holevo} of $B$,
$R(B)$  may have a  maximum value  less than  $\log(d)$ in  the finite
dimensional  case.   A  generalization  of  the  concept  of  `maximal
complementarity' or `MUBness' would be to apply those terms to $A$ and
$B$, when  one of them is a  POVM, where the maximal  knowledge of the
measured value of $A$ implies  minimal knowledge of the measured value
of $B$,  and vice  versa, but with  maximum knowledge no  longer being
required to be as high as $\log d$ bits.

For  the phase  variable  given  by the  POVM  $\phi$ and  probability
distribution  ${\cal P}(\phi)$,  entropic  knowledge is  given by  the
functional \cite{sb06,sr07}:
\begin{equation}
R[{\cal P}(\phi)] = \int_0^{2\pi}d\phi~
{\cal P}(\phi)\log[2\pi {\cal P}(\phi)],
\label{eq:phient}
\end{equation}
where the $\log(\cdot)$ refers to the binary base.

It is at first not  obvious that Eq.  (\ref{eq:ra}) holds for infinite
dimensional systems.   Based on a result due  to Bialynicki-Birula and
Mycielski \cite{bial}, which  in turn uses the concept of
the  $(p,q)$-norm  of  the  Fourier transformation  found  by  Beckner
\cite{beck75} for  all values of $p$, for an oscillator
system, we can show that it is indeed the case.  In particular,
\begin{equation}
\label{eq:bial}
-\int_{-\pi}^\pi   d\phi   P(\phi)\log(P(\phi))  -   \sum_{m=0}^\infty
p_m\log(p_m) \ge \log(2\pi)
\end{equation}
Here it is worth
noting that, along the lines of Ref. \cite{bial}, 
one may obtain analogous entropic uncertainty relations between phase and
number of quanta, as well as between energy and time \cite{gra87}.

Setting the `number variable' $m$ in Eq. (\ref{eq:bial}) as $A$, and the
phase variable $\phi$ as $B$, and noting that
the first term in the l.h.s of Eq. (\ref{eq:bial}), 
using Eq. (\ref{eq:phient}), is just $\log(2\pi) -
R[P(\phi)]$, we obtain
\begin{equation}
X[m, \phi] \equiv H[m] - R[\phi] \ge 0,
\label{eq:nybial}
\end{equation}
which is  Eq. (\ref{eq:ra}) applied to  an infinite-dimensional system
that   includes   a    non-Hermitian   POVM   (phase   $\phi$).    Eq.
(\ref{eq:nybial}) expresses  the fact ignorance of variable  $m$ is at
least  as great  as knowledge  of its  complementary  partner, $\phi$.
Restricting our attention only to number-phase complementarity,
we find on
comparing Eqs.  (\ref{eq:ra}) and  (\ref{eq:nybial}) that the
statement $X  \ge 0$  as a description  of the  Heisenberg uncertainty
relation 
holds  good both for  finite and  infinite dimensional  systems.  This
version of the Heisenberg uncertainty principle may be called {\it the
  principle  of  entropy  excess}. This  thus renders physically
intuitive the result of Bialynicki-Birula and Mycielski \cite{bial},
derived  using   elements  of  advanced  functional   analysis.
As related work, we cite an
information theoretic interpretation of  uncertainty in the context of
phase  resolution  in  harmonic  oscillator systems,  in  \cite{mh93}.
Also,  the  number-phase complementarity,  for  a harmonic  oscillator
system,  using information  exclusion  relations has  been studied  in
\cite{mh95}.


As pointed  out earlier, the knowledge-sum approach  cannot be applied
to  infinite dimensional  systems,  whereas the  principle of  entropy
excess  can be  applied  to  finite as  well  as infinite  dimensional
systems, making  it a more  flexible tool for  describing number-phase
complementarity in a host of  systems.  Here we apply the principle of
entropy  excess (\ref{eq:nybial})  to number-phase  complementarity in
(finite-level)  atomic systems,  briefly  revisiting results  obtained
earlier  \cite{rs07} from  the perspective  of an  upper bound  on the
knowledge-sum of  complementary variables, as  well as to  an infinite
dimensional  harmonic  oscillator,  thereby highlighting  its  greater
scope.

When  applied  to  a   finite  level  (atomic)  system,  the  relation
(\ref{eq:nybial}) still leaves  some room for improvement \cite{rs07}.
For example, in the case  of qubits (two-level systems), number states
saturate  the   bound  because  they   satisfy  $H(m)=R(\phi)=0$.  The
corresponding  states  of the  phase  variable  (which maximize  phase
knowledge and  minimize number  knowledge) are the  equatorial states,
for  which $H(m)=1$,  but  $R[\phi] \approx  0.245  < 1$  \cite{rs07}.
Following this reference, one way to address this problem is to modify
(\ref{eq:nybial}) to the inequality
\begin{equation}
\label{eq:2bit}
X^\mu[m,\phi] \equiv H[m] - \mu R[\phi] \ge 0
\end{equation}
for all  pure states in ${\bf  C}^2$, where parameter  $\mu$ $(>0)$ is
chosen to be the largest value such that inequality (\ref{eq:2bit}) is
satisfied over all state space. 

\begin{figure}
\subfigure[]{\includegraphics[width=7.0cm]{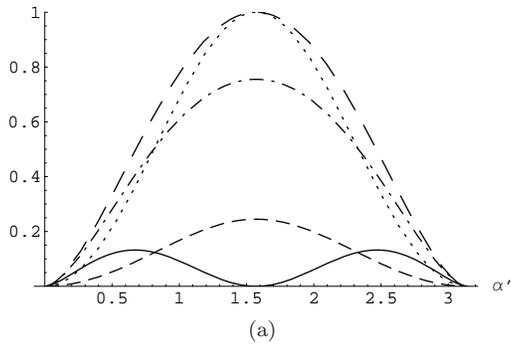}}
\caption{Entropy excess  of a two  level atomic system starting  in an
  atomic  coherent  state  $|\alpha^\prime,\beta^\prime\rangle$, as  a
  function of $\alpha^\prime$, with $\beta^\prime=0.0$, two parameters
  covering the Bloch  sphere of a two-level system  in the notation of
  Ref.  \cite{rs07}.   The  large-dashed  (resp.,  small-dashed)  line
  represents $H[m]$  (resp., $R[\phi]$).  The  dotted-curve represents
  $\mu  R[\phi]$ (where $\mu=4.085$).   The solid  (resp., dot-dashed)
  curve represents the entropy excess $X_\mu$ (resp.  $X$).}
\label{fig:atomXS}
\end{figure}

\begin{figure}
\subfigure[]{\includegraphics[width=7.0cm]{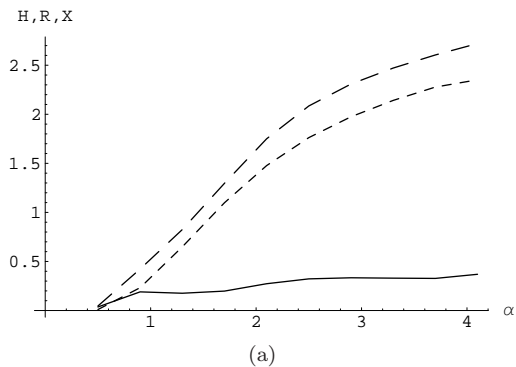}}
\caption{ Number  entropy $H[m]$ (large-dashed  line), phase knowledge
  $R[\phi]$ (small-dashed  line) and entropy excess  $X[m, \phi]$ (Eq.
  (\ref{eq:nybial}), bold line) plotted as a function of the parameter
  $\alpha$ for  a harmonic oscillator  system initially in  a coherent
  state $|\alpha \rangle$.}
\label{fig:inicoh}
\end{figure}

From  the  concavity  of $H[m]$  and  the  convexity  of
$R[\phi]$,  it follows that  Eq. (\ref{eq:2bit})  holds for  any mixed
state.   Figure  (\ref{fig:atomXS}) ,  illustrates  the tighter  bound
imposed by $X_\mu[m,\phi]$ than $X[m,\phi]$.
In  Figure  (\ref{fig:inicoh}),   the  number  entropy  $H[m]$,  phase
knowledge   $R[\phi]$   and   entropy   excess   $X[m,   \phi]$   (Eq.
(\ref{eq:nybial})) are depicted for a harmonic oscillator starting out
in the usual coherent state  $|\alpha\rangle$.  

We note that as number
increases,  with  increase in  $\alpha$,  so  does  $H[m]$ (since  the
variance  of a Poisson  distribution equals  its mean),  whereas phase
$\phi$ becomes increasingly certain, leading to increase in $R[\phi]$.
Through a numerical search, we found
that $\mu \approx  4.085$ for dimension $d=2$ and  $\mu \approx 1.973$
for  $d=4$. And when $d = \infty$, we find analytically $\mu = 1$,   
as  can be seen from  the discussion leading up to  Eq.  (\ref{eq:nybial}).  
From
the  above numerical-analytical  pattern,  we conjecture  that as  the
system  dimension   increases  from  two  to   infinity,  $\mu$  falls
monotonically from about 4 to 1.

Thus  the principle  of  entropy excess,  incorporating knowledge  and
entropy,  emerges   as  a  flexible  measure   by  which 
number-phase complementarity  of  finite  as  well  as
infinite dimensional systems can be studied in a unified manner.

\end{document}